\documentclass[a4paper]{spie}  

\usepackage{amsmath,amsfonts,amssymb}
\usepackage{graphicx, subcaption}
\usepackage{tabularx}
\usepackage[colorlinks=true, allcolors=blue]{hyperref}
\usepackage{xcolor}
\usepackage[normalem]{ulem}
\usepackage{caption}
\usepackage{multirow}
\usepackage{csquotes}

\title{Mixed-Block Neural Architecture Search for Medical Image Segmentation}

\author[a, b]{Martijn M.A. Bosma}\author[a]{Arkadiy Dushatskiy\textsuperscript{1}}
\author[a]{Monika Grewal\thanks{\hspace{0.3ex} equal contribution}\hspace{2mm}}
\author[c]{Tanja Alderliesten}
\author[a, b]{Peter A. N. Bosman}
\affil[a]{Centrum Wiskunde \& Informatica, Science Park 123, 1098XG, Amsterdam, the Netherlands}
\affil[b]{Delft University of Technology, Mekelweg 5, 2628CD, Delft, the Netherlands}
\affil[c]{Leiden University Medical Center, Albinusdreef 2, 2333ZA, Leiden, the Netherlands}

\begin{document} 
\maketitle

\begin{abstract}
Deep Neural Networks (DNNs) have the potential for making various clinical procedures more time-efficient by automating medical image segmentation. Due to their strong, in some cases human-level, performance, they have become the standard approach in this field. The design of the best possible medical image segmentation DNNs, however, is task-specific. Neural Architecture Search (NAS), i.e., the automation of neural network design, has been shown to have the capability to outperform manually designed networks for various tasks. However, the existing NAS methods for medical image segmentation have explored a quite limited range of types of DNN architectures that can be discovered. In this work, we propose a novel NAS search space for medical image segmentation networks. This search space combines the strength of a generalised encoder-decoder structure, well known from U-Net, with network blocks that have proven to have a strong performance in image classification tasks. The search is performed by looking for the best topology of multiple cells simultaneously with the configuration of each cell within, allowing for interactions between topology and cell-level attributes. From experiments on two publicly available datasets, we find that the networks discovered by our proposed NAS method have better performance than well-known handcrafted segmentation networks, and outperform networks found with other NAS approaches that perform only topology search, and topology-level search followed by cell-level search. 
\end{abstract}


\section{INTRODUCTION}
\label{sec:intro}  
A growing amount of clinical applications, such as computer-aided diagnostic systems, are benefiting from recent advances in automated medical image segmentation, most notably from Deep Neural Networks (DNNs) \cite{CARDIAC_DIAG, RETINAL_DIAG}. Given a medical scan, a DNN can provide contours of organs or regions of interest (e.g., tumors), with clinically acceptable segmentation quality within a matter of seconds \cite{surfacedice}. Designing a State-of-the-Art (SotA) DNN, however, is often task-specific. In order to design a DNN, choices have to be made for the topology of the network such as depth, and connections between cells (a cell is a group of operations that transform the feature maps in a DNN), as well as the configuration of each cell, e.g., convolutional kernel size, or activation function. This gives rise to an inconceivably large amount of network architecture design possibilities, which is impossible to manually navigate through in an exhaustive fashion, or even by means of intelligent design, while ensuring the best choices are made.

Neural Architecture Search (NAS), i.e., the automated design of neural network architectures, can effectively and efficiently search through this space of possible network architecture designs and find a network that is highly tailored to the task at hand\cite{NAS_SURVEY}. While research on NAS for medical image segmentation has not been as elaborate as for natural image classification, it has already shown promising results by outperforming the SotA architectures\cite{C2FNAS, MSNAS, NASUNET}. In our opinion, further research on NAS for medical image segmentation can make its contributions even more significant.

NAS involves three key components: (1) The search space (the set of all possible networks given the specified architectural constraints); (2) the search algorithm (the algorithm to navigate the search space); (3) performance estimation (the choices made to score a network's performance, such that these networks can be ranked by the search algorithm). So far, NAS research has been more elaborate for image classification tasks\cite{NAS_SURVEY}. Potentially, strong performing search algorithms, and fast and accurate performance estimation methods can be translated to the medical image segmentation domain. The search space, however, is specific to segmentation tasks. Several papers have proposed search spaces for medical image segmentation. These include searching for a U-Net like encoder-decoder topology i.e., by adapting only the cells within\cite{NASUNET}, by searching for the best topology followed by the best convolution size within cells\cite{C2FNAS}, or a combination of topology and cell search using continuous relaxation and a so-called Super-network \cite{MSNAS}. With each of these NAS search spaces it has been possible to find networks that perform slightly better than a standard U-Net. A semi-automatic network architecture design and training method known as nnU-Net \cite{NNUNET}, however, is still considered SotA for many image segmentation tasks.  

We believe that within the scope of a U-Net-like topology search, there is a room for making the search space more flexible -- instead of repeating the same cell structure, we propose to allow various cell structures in a network, potentially resulting in a better performance. Further, we believe that the configuration of cells can benefit from existing knowledge about networks with a strong performance on classification tasks. Therefore, we propose to search through a pre-selected pool of configurations (which are taken from advanced well-known classification networks \cite{RESNET, DENSENET, INCEPTIONNET}) instead of searching for a cell configuration from scratch. Apart from benefiting from advanced performance of these well known classification networks, this proposed improvement will also help avoiding the explosive growth of the search space caused by searching the configuration of each cell from scratch. Finally, in contrast to recent research \cite{C2FNAS}, where topology level search is followed by cell level search, we search for both topology as well as the configuration of each cell simultaneously. This allows to take into consideration the possible interaction between the topology-level search and cell-level, potentially yielding better performing networks. The combination of these improvements results in the proposed approach, which we will further refer to as Mixed-Block NAS (\textbf{MB-NAS}).

\section{METHOD}
\label{sec:method}
\subsection{Search space}
\label{sec:searchspace}

\begin{figure} [ht]
    \begin{center}
    \begin{subfigure}[t]{\linewidth}
    \centering\includegraphics[width=\linewidth]{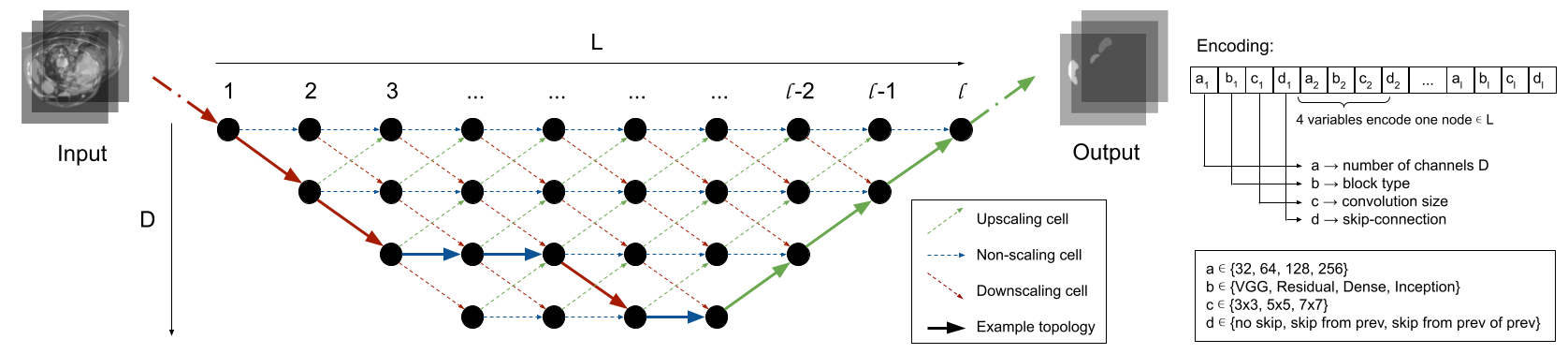} 
    \caption{\scriptsize Topology-level search and encoding.}
    \label{subfig:search_macro}
    \end{subfigure} \\
    \begin{subfigure}[t]{.41\linewidth}
    \centering\includegraphics[width=\linewidth]{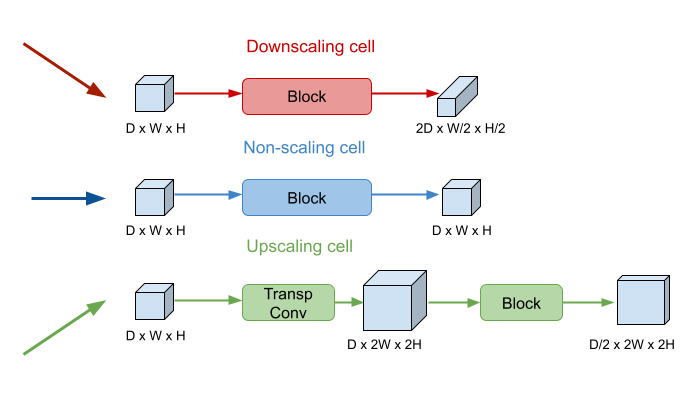}
    \caption{\scriptsize Cell types.}
    \label{subfig:search_cells}
    \end{subfigure}
    \begin{subfigure}[t]{0.58\linewidth}
    \centering\includegraphics[width=\linewidth]{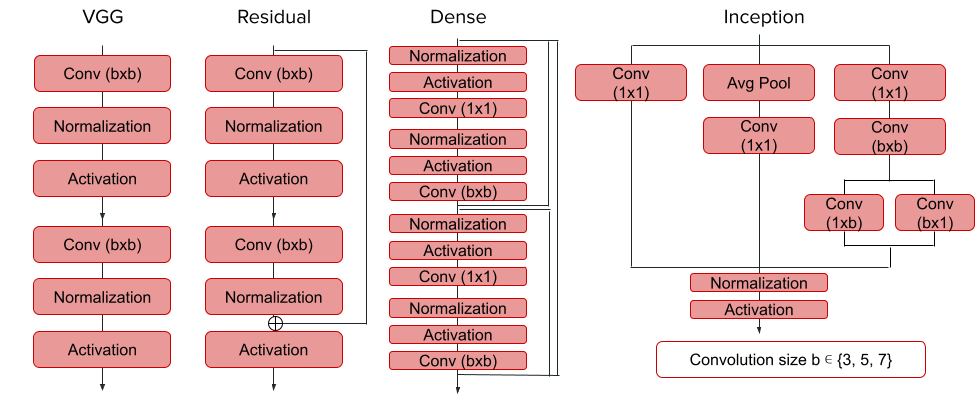}
    \caption{\scriptsize Block types.}
    \label{subfig:search_blocks}
    \end{subfigure}
    \end{center}
    \caption[example] 
    { \label{fig:space}
    \small Description of MB-NAS search space.}
\end{figure}
At \textbf{topology-level} (Figure \ref{subfig:search_macro}), the search space contains all possible network architectures resulting from varying connections between cells of different types. We consider three possibilities for a cell (Figure \ref{subfig:search_cells}): a downscaling, an upscaling, or a non-scaling cell. A downscaling cell means that the input image resolution is halved, while the number of channels is doubled. Upscaling is the opposite operation, i.e., doubling the resolution and halving the number of channels, while a non-scaling cell changes neither the resolution nor the number of channels in the feature maps. The input to a cell can be either just the feature maps from the preceding cell (no skip-connection), or the feature maps of the preceding cells concatenated with the feature maps from any previous cell with the same feature maps spatial dimensions (a single skip-connection).

At \textbf{cell-level}, different configurations of each cell are searched. The configuration of a cell is encoded by two variables: the type of block (a block is an organised structure consisting of multiple convolution and normalisation layers, as well as activation functions) within the cell, and the convolutional kernel size within the block. Instead of searching for the topology of the blocks within a cell, which would make the search space incredibly large, we used predefined blocks derived from previously SotA architectures for classification (Figure \ref{subfig:search_blocks}). In this work we consider VGG blocks \cite{VGG} which are standard in U-Net, as well as Residual blocks \cite{RESNET}, Dense blocks \cite{DENSENET} and Inception blocks \cite{INCEPTIONNET}. In this way, we allow the search space to chose a different cell configuration at every edge (instead of repeating one throughout the network) while preventing the further growth of the search space.

The network architecture is represented by connections between a fixed number of nodes (fixed to 10 in our experiments). Each node $l$ (Figure \ref{subfig:search_macro}) is represented by 4 categorical variables: $a_l$ = number of channels, $b_l$ = the block type, $c_l$ = convolution size, and $d_l$ = skip-connection source. The cell type is derived by the difference in number of channels between two nodes. The topology of the neural network is encoded by variables $a_l$ and $d_l$ at each node. Variable $a_l$ is restrained to only be allowed to take on double, equal or half the number of channels contained in the previous cell. The cell-level search is represented by variables $b_l$ and $c_l$ at each node. Note that the standard U-Net shape is included in the topology-level search space. The resulting search space contains $1.14*10^{18}$ possible networks.

\subsection{Datasets}
In our experiments, we used two datasets that can be found on the Medical Decathlon\cite{MSD} challenge website\footnote{http://medicaldecathlon.com/}. The first is a collection of 3D MRI-scans of the prostate, alongside their multi-class (Central and Peripheral prostate zones) segmentations. It contains 32 multi-modal scans, represented as a collection of 602 2D slices total. The scans were normalised, using Z-normalisation. The second contains 3D-CT scans and single class segmentations of the spleen. It contains 41 single modality scans represented as 3650 2D slices. The scans were clipped and normalised. 

\subsection{Implementation details}
\label{sec:searchalgos}
We use Local Search (LS) as the search algorithm as it has been shown to be a strong baseline for NAS\cite{LSSTRONG}. Specifically, we use a first-improvement approach with a variable neighbourhood of 1. This means that for each variable (iterated in a random order), all possible options are probed, and the best performing is chosen. Once all variables are considered, we start over, until no improvements can be found anywhere. 
Due to limited computational resources, the NAS was run for only 150 network evaluations. Each network was evaluated using the validation Dice score averaged over the last 20\% of training epochs. The number of epochs (100 epochs for Prostate, 50 for Spleen) was decided such that saturation was ensured based on preliminary runs. We used the average score of 5-fold patient-level cross-validation, repeated for 3 different network initialisations, as performance estimator. Multiple initialisations reduce the noise in scoring, providing more reliable information to the search algorithm, making it easier to navigate the search space. The use of multiple folds decreases noise caused by data splits. 

For network training, we used the ADAM optimiser with a learning rate $10^{-3}$ and polynomial decay with an exponent of $0.9$. The loss function was foreground soft Dice. The batch size was 32 and input image size was 128x128. The data was augmented using scaling, shifting, rotating, flipping, and brightness adjustment to avoid overfitting.

\subsection{Experimental setup}
\label{sec:data}
We performed experiments to compare the proposed search space design (Mixed-Block, or \textbf{MB-NAS}) against the following alternative approaches. In the first considered alternative search space (\textbf{Macro-NAS}) only topology search is performed. All block types are fixed to be standard U-Net blocks (VGG). The second considered search space (\textbf{Micro-NAS}) has the U-Net topology, where only the block type and convolution sizes are subject to the search. In the third alternative search space (\textbf{Bilevel-NAS}), a bi-level approach was used which means that first the topology-level is searched, and then the convolution size, similar to the Coarse-to-Fine approach\cite{C2FNAS}.

The best networks found by the proposed search space were evaluated against two hand-crafted neural network architectures: standard \textbf{U-Net}, and U-Net with a ResNet-50 encoder (\textbf{ResU-Net}). Implementations are taken from the Pytorch Segmentation Models library \cite{SMP}.


\section{RESULTS AND DISCUSSION}
\begin{figure} [ht]
   \begin{center}
    \begin{subfigure}[t]{.3\linewidth}
    \centering\includegraphics[width=\linewidth]{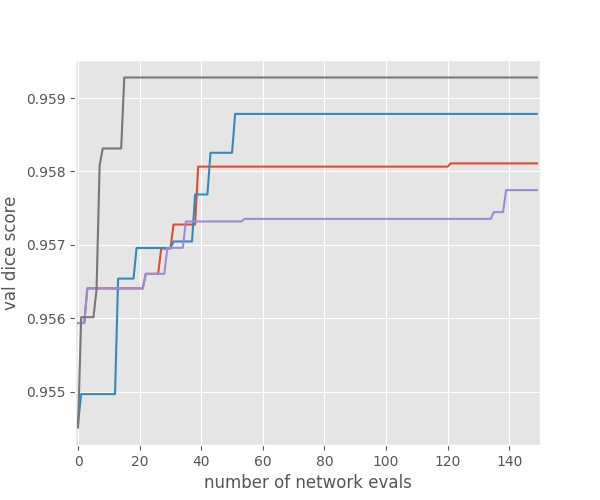}
    \caption{\scriptsize NAS Spleen dataset.}
    \label{subfig:nas_spleen}
    \end{subfigure}
    \begin{subfigure}[t]{.3\linewidth}
    \centering\includegraphics[width=\linewidth]{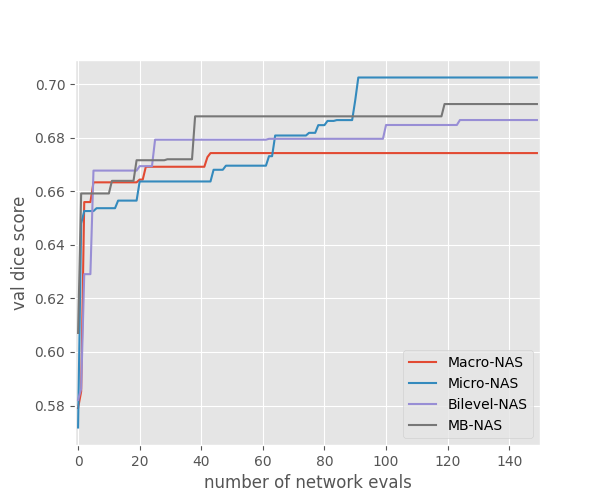}
    \caption{\scriptsize NAS Prostate dataset.}
    \label{subfig:nas_prostate}
    \end{subfigure}
    \begin{subfigure}[t]{.33\linewidth}
    \centering\includegraphics[width=\linewidth]{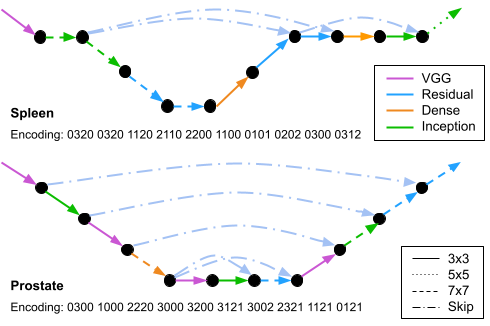}
    \caption{\scriptsize Best networks from MB-NAS.}
    \label{subfig:elites}
    \end{subfigure} \\
    \begin{subfigure}[t]{.49\linewidth}
    \centering\includegraphics[width=\linewidth]{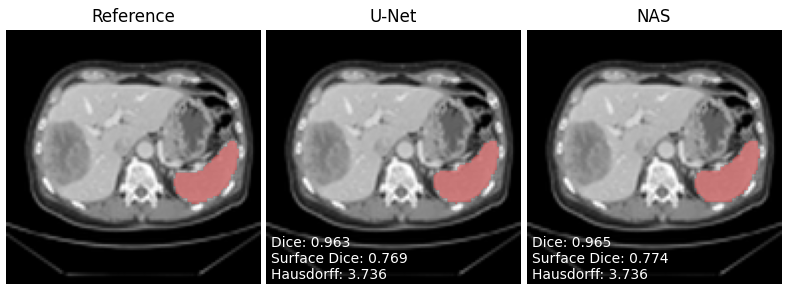}
    \caption{\scriptsize Example segmentations on the Spleen dataset.}
    \label{subfig:segs_spleen}
    \end{subfigure}
    \begin{subfigure}[t]{.49\linewidth}
    \centering\includegraphics[width=\linewidth]{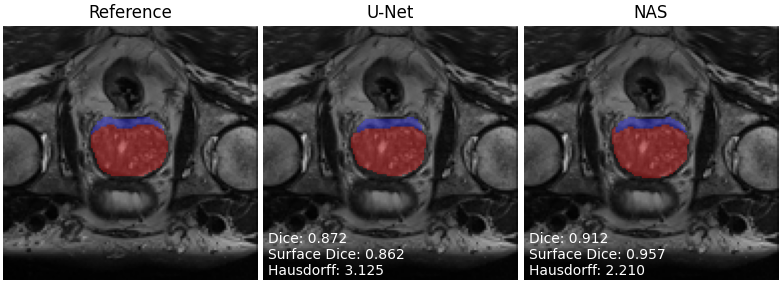}
    \caption{\scriptsize Example segmentations on the Prostate dataset. red: central gland, purple: peripheral zone.}
    \label{subfig:segs_prostate}
    \end{subfigure}
   \end{center}
   \caption[.] 
   { \label{fig:nas}
   \small The progress of the LS algorithm for the (a) Spleen and (b) Prostate datasets, respectively. The best networks found by LS with MB-NAS are shown in (c). (d) and (e) show example segmentations.}
\end{figure}

The performance of the different search spaces (\textbf{MB-NAS}, \textbf{Macro-NAS}, \textbf{Micro-NAS}, \textbf{Bilevel-NAS}) can be seen in Figure \ref{subfig:nas_spleen}  and \ref{subfig:nas_prostate}. The the architectures of the best networks from MB-NAS are visualised in Figure \ref{subfig:elites}. The performance of the best networks from different NAS approaches, \textbf{U-Net}, and \textbf{ResU-Net} is summarised in Table \ref{tab:results}. 

\begin{table}[ht]
    \caption[Performance comparison of MB-NAS]{Performance values of U-Net, ResU-Net, and the best networks from different NAS approaches. The values are averaged over 5 seeds on a 5-fold cross-validation. Standard deviations are calculated based on single seed scores on all 5-folds. Best values in each column are highlighted in bold. DSC: Dice-Sorensen similarity coefficient, HD: Hausdorff distance (95\% cutoff), SD: Surface Dice (2mm threshold), MMAC: Mega Multiply–ACcumulate operations, \#Params: number of parameters, $\times 10^6$. } 
    \label{tab:results}
    \begin{center}      
    \scalebox{0.95}{
    \begin{tabular}{|l|l|l|l|l|l|}
    \hline
    \multirow{2}{6em}{Model} & \multicolumn{5}{c|}{Prostate dataset} \\
    \cline{2-6}
    & & & & & \\[-1em]
    \rule[-1ex]{0pt}{1ex} & DSC & HD & SD & MMAC & \#Params \\
    \hline
    \rule[-1ex]{-1pt}{1.0ex}  \textbf{U-Net} & $0.6702 \pm 0.004$ & $8.833 \pm 0.348$ & $0.6046 \pm 0.006$ & $302$ & $18.4$  \\
    \rule[-1ex]{-1pt}{1.0ex}  \textbf{ResU-Net} & $0.6580 \pm 0.004$ & $9.441 \pm 0.325$ & $0.5705 \pm 0.006$ & $\mathbf{166}$ & $32.5$  \\
    \rule[-1ex]{-1pt}{1.0ex}  \textbf{Macro-NAS} & $0.6593 \pm 0.004$ &  $8.606 \pm 0.470$ & $0.5977 \pm 0.008 $ & $256$ & $3.39$ \\
    \rule[-1ex]{-1pt}{1.0ex}  \textbf{Micro-NAS} & $\mathbf{0.6796\pm 0.007}$ & $\mathbf{8.394 \pm 0.251}$ & $\mathbf{0.6203 \pm 0.008}$ & $1,295$ & $22.7$\\
    \rule[-1ex]{-1pt}{1.0ex}  \textbf{Bilevel-NAS} & $0.6702 \pm 0.005$ & $8.492 \pm 0.541$ & $0.6134 \pm 0.008 $ & $414$ & $6.77$ \\
    \rule[-1ex]{-1pt}{1.0ex}  \textbf{MB-NAS} & $0.6760 \pm 0.010$ & $8.419 \pm 0.324$ & $0.6192 \pm 0.004$ & $644$ & $\mathbf{3.04}$ \\
    \hline 
    \end{tabular} 
    }
    \end{center}
    \begin{center}      
    \scalebox{0.95}{
    \begin{tabular}{|l|l|l|l|l|l|l|l|l|l|l|}
    \hline
    \multirow{2}{6em}{Model} &  \multicolumn{5}{c|}{Spleen dataset} \\
    \cline{2-6}
    & & & & & \\[-1em]
    \rule[-1ex]{0pt}{1ex} & DSC & HD & SD & MMAC & \#Params \\
    \hline
    \rule[-1ex]{-1pt}{1.0ex}  \textbf{U-Net} & $0.9578 \pm 0.002$ & $1.412 \pm 0.030$ & $0.917 \pm 0.002$ & $302$ & $18.4$  \\
    \rule[-1ex]{-1pt}{1.0ex}  \textbf{ResU-Net} & $0.9464 \pm 0.004$ & $1.625 \pm 0.042$ & $0.905 \pm 0.003$ & $\mathbf{166}$ & $32.5$ \\
    \rule[-1ex]{-1pt}{1.0ex}  \textbf{Macro-NAS} & $0.9566 \pm 0.001$ &  $1.467 \pm 0.063$ & $0.9167 \pm 0.001$ & $255$  & $\mathbf{3.39}$ \\
    \rule[-1ex]{-1pt}{1.0ex}  \textbf{Micro-NAS} & $0.9567 \pm 0.002$ & $1.388 \pm 0.026$ & $0.9177 \pm 0.004$ & $795$  & $2.83$ \\
    \rule[-1ex]{-1pt}{1.0ex}  \textbf{Bilevel-NAS} & $0.9553 \pm 0.001$ & $1.449 \pm 0.069$ & $0.9145 \pm 0.005$ & $415$ & $6.78$  \\
    \rule[-1ex]{-1pt}{1.0ex}  \textbf{MB-NAS} & $\mathbf{0.9592 \pm 0.002}$ & $\mathbf{1.385 \pm 0.036}$ & $\mathbf{0.9189 \pm 0.002}$ & $1,294$ & $22.7$ \\
    \hline 
    \end{tabular}
    }
    \end{center}
\end{table}

\subsection{Spleen dataset}
The best performing network on the Spleen dataset was found using the proposed search space, MB-NAS. In Table \ref{tab:results}, one can see that this network shows the best performance by all three considered performance metrics. Note that the topology (Figure~\ref{subfig:elites}, top row) is quite different from the standard U-Net. The network is shallower, potentially indicating that smaller, more intricate features, are more important in the segmentation of the spleen, than larger, more elaborate ones. This could be due to the small size of the organ in relation to the image. Furthermore, blocks of all four types, are included in the architecture, as well as different convolution sizes. This means that both the topology-level and cell-level search space parts were utilised to find this network; an indication towards the performance increase from the designed search space.

\subsection{Prostate dataset}
Different than for the Spleen dataset, the best network (Figure \ref{subfig:elites}, bottom row) is very similar to the U-Net architecture. This provides additional evidence to the argument that the architecture of the best DNN is task-specific. This also indicates that the U-Net topology is best suited for the underlying task, giving an added advantage to Micro-NAS, wherein different blocks and convolution sizes are searched within the U-Net topology. Consequently, the use of Micro-NAS yields the best performance. However, it should be noted that using MB-NAS, we could find a network with comparable performance despite the fact that the search was performed for both topology as well as cell configuration. 
It is worth noting that no block seems to be ultimately preferable to other blocks, i.e. outperforms other blocks at every location of the network. This indicates the advantage of searching for a mixed-block configuration such that different blocks can be used throughout the network. Additionally, the differences in the validation DSC of poorly performing networks compared to the best networks is much larger for the Prostate dataset, indicating a more difficult task. Further, Table \ref{tab:results} shows that the best networks from NAS outperform the manually hand-crafted U-Net and ResU-Net networks.



\subsection{Comparison with SotA}
It can be argued that the results for the Prostate dataset in Table \ref{tab:results} are not at the same level as the SotA results given by e.g., nnU-Net \cite{NNUNET}. This is due to higher image resolution used by nnU-Net, additional preprocessing, carefully chosen data augmentations, post-processing, and an advanced inference method, that are all different from the setup used in NAS in this paper. Therefore, the network performance when using the nnU-Net training and evaluation setup, was also evaluated with the found networks from NAS. For Prostate, the 5-fold cross-validation Dice-Sorensen coefficient is $\mathbf{0.7325}$ for MB-NAS vs. $\mathbf{0.7315}$ for nnU-Net. It should be noted that the comparison is done between the network found by NAS, trained in the nnU-Net environment, and a U-Net architecture that is tailored to the data according to the nnU-Net heuristics. The architectures found by NAS were found using entirely different settings. This gives an unfair advantage to nnU-Net. Nevertheless, the architecture found by NAS still performs slightly better to nnU-Net in these settings, which is remarkable. For the Spleen dataset, the performance level is quite similar to the results reported by nnU-Net. Nevertheless, the network was validated in this environment as well. The 5-fold cross-validation Dice-Sorensen coefficient is $\mathbf{0.9467}$ for MB-NAS vs. $\mathbf{0.9466}$ for nnU-Net, indicating equal performance. 


\subsection{Statistical significance}


A Friedman test was conducted on the different network performances to compare the effect of the architecture on the segmentation accuracy. The test was performed on individual scores for 5 differently seeded network initialisations on all 5 folds of a datasplit. The Friedman test reveals a statistically significant difference between network performances from Table \ref{tab:results} for the Dice-Sorensen similarity coefficient (Spleen dataset : $(\chi^2(5) = 49.14, p<0.0001)$; Prostate dataset : $(\chi^2(5) = 16.98, p<0.005)$). A post-hoc analysis, using the Wilcoxon signed-rank test, found that the difference in performance of the Dice-Sorensen similarity coefficient was statistically significant between the best network found by NAS and U-Net for the Spleen dataset ($Z=48.0, p<0.05$). For the Prostate dataset the difference was not significant.

\subsection{Computational limits and possible extensions}
We note that due to limited computational resources, only one run of NAS was performed per search space, and for a limited number of network evaluations. Longer experiments with multiple runs would have helped draw more definitive conclusions. Also, running NAS experiments for more datasets may provide further insights about the specific characteristics of network architecture design required for good performance across datasets. Additionally, the search space could be further extended by adding other block types, like Image Transformer \cite{TRANSFORMERS} blocks.

\section{Conclusions}
\label{sec:conclusions}
We have proposed an approach for medical image segmentation Neural Architecture Search, which involves a novel search space and simultaneous topology- and cell-level search strategy. In the cell-level search, we used existing knowledge from networks with high performance in image classification tasks, i.e. ResNet, DenseNet and InceptionNet, to create a pool of possible block configurations. The experiments in this paper show the added value of this approach. Overall, the results indicate that further research into search space refinement, allowing to exploit key features of what accounts for good deep learning performance, may yet push the boundaries of what can be achieved with deep neural networks for medical image processing.

\bibliography{report} 
\bibliographystyle{spiebib} 

\end{document}